\documentclass[preprint,showpacs,preprintnumbers,amsmath,amssymb]{revtex4}
\usepackage{graphicx}
\usepackage{dcolumn}
\usepackage{bm}

\def\lesssim{\ \raise.3ex\hbox{$<$}\kern-0.8em\lower.7ex\hbox{$\sim$}\ }
\def\gesim{\ \raise.3ex\hbox{$>$}\kern-0.8em\lower.7ex\hbox{$\sim$}\ }

\def\lesssim{\ \raise.3ex\hbox{$<$}\kern-0.8em\lower.7ex\hbox{$\sim$}\ }
\def\gesim{\ \raise.3ex\hbox{$>$}\kern-0.8em\lower.7ex\hbox{$\sim$}\ }
\font\scripti=cmmi7
\font\scriptscripti=cmmi5
\def\sib#1{\setbox0 = \hbox{\scripti #1}
  \kern-.02em\copy0\kern-\wd0
  \kern.04em\box0} 
\def\ssib#1{\setbox0 = \hbox{\scriptscripti #1}
  \kern-.02em\copy0\kern-\wd0
  \kern.04em\box0} 
\font\tenib=cmmib10 
\skewchar\tenib='177 \skewchar\tenib='177 \skewchar\tenib='177
\def\pbold#1{\setbox0 = \hbox{$ #1 $}
  \kern-.022em\copy0\kern-\wd0
  \kern.011em\copy0\kern-\wd0
  \kern.011em\copy0\kern-\wd0
  \kern.011em\copy0\kern-\wd0
  \kern.011em\box0} 

\usepackage{graphicx}
\usepackage{dcolumn}
\usepackage{bm}

\def\ev#1{\langle #1 \rangle}
\def\cp{\mu}

\def\cpdwn{\mu_\downarrow}
\def\op{\Delta}
\def\s{\sigma}
\def\up{\uparrow}
\def\dwn{\downarrow}

\begin{document}

\title{Superfluid/ferromagnet/superfluid junction and $\pi$-phase in a superfluid Fermi gas}

\author{Takashi Kashimura$^{1}$, Shunji Tsuchiya$^{2,3}$, and Yoji Ohashi$^{1,3}$}\affiliation{
$^1$Department of Physics, Keio University, 3-14-1 Hiyoshi, Kohoku-ku, Yokohama 223-8522, Japan,\\
$^2$Department of Physics, Tokyo University of Science, 1-3 Kagurazaka, Shinjuku-ku, Tokyo 162-8601, Japan,\\
$^3$CREST(JST), 4-1-8 Honcho, Saitama 332-0012, Japan} 

\date{\today}

\begin{abstract}
We investigate the possibility of superfluid/ferromagnet/superfluid (SFS)-junction in a superfluid Fermi gas. To examine this possibility in a simple manner, we consider an attractive Hubbard model at $T=0$ within the mean-field theory. When a potential barrier is embedded in a superfluid Fermi gas with population imbalance ($N_\uparrow>N_\downarrow$, where $N_\sigma$ is the number of atoms with pseudospin $\sigma=\uparrow,\downarrow$), this barrier is shown to be {\it magnetized} in the sense that excess $\uparrow$-spin atoms are localized around it. The resulting superfluid Fermi gas is spatially divided into two by this {\it ferromagnet}, so that one obtains a junction similar to the superconductor/ferromagnet/superconductor-junction discussed in superconductivity. Indeed, we show that the so-called $\pi$-phase, which is a typical phenomenon in the SFS-junction, is realized, where the superfluid order parameter changes its sign across the junction. Our results would be useful for the study of magnetic effects on fermion superfluidity using an ultracold Fermi gas. 

\end{abstract}

\maketitle

\section{Introduction} \label{intro}
The advantage of recently realized superfluid $^{40}$K and $^6$Li Fermi gases\cite{Regal,Bartenstein,Zwierlein,Kinast} is that one can experimentally tune various physical parameters. This unique property enables us to study various physical properties of Fermi superfluids in a wide parameter region\cite{Chen}. Since the background physics of superfluid Fermi gas is essentially the same as superconductivity, the study of former atomic system is also useful for understanding the latter electron system. Indeed, using a tunable interaction associated with a Feshbach resonance\cite{Regal,Bartenstein,Zwierlein,Kinast,Chin}, one can study superfluid properties from the weak-coupling regime to the strong-coupling limit in a unified manner\cite{Eagles,Leggett,NSR,SadeMelo,Holland,Ohashi}. In the intermediate coupling regime (crossover region), strong pairing fluctuations lead to the pseudogap phenomenon\cite{Jin1,Jin2,Tsuchiya1,Tsuchiya2,HuiHu}, which is related to the under-doped regime of high-$T_{\rm c}$ cuprates\cite{Lee,Randeria,Singer,Janko,Rohe,Yanase,Perali}. In addition, a cold Fermi gas loaded onto an optical lattice is well described by the Hubbard model\cite{Stringari,Kohl,MIT1,MIT2}, which is one of the most fundamental models in strongly-correlated electron system. Thus, using this lattice Fermi gas, we may study various phenomena originating from strong correlation, such as Mott insulator\cite{Esslinger1}, and unconventional pairing state\cite{Demler}. We also point out that a superfluid Fermi gas with population imbalance is similar to superconductivity under an external magnetic field, so that the possibility of Fulde-Ferrel-Larkin-Ovchinnikov(FFLO) state\cite{FF,LO,Takada} has been recently discussed in polarized Fermi gases\cite{Castorina,Mizushima,Yang,Kinnunen,Machida}. 
\par
The above examples clearly show the importance of cold Fermi gas as a {\it quantum simulator} for metallic superconductivity. However, at the same time, we also know that the current stage of superfluid Fermi gas cannot cover all the important topics discussed in superconductivity. In metallic superconductivity, magnetism is known to strongly affect superconducting properties\cite{DeGennes,Maki}. In particular, the conventional $s$-wave pairing is easily destroyed by ferromagnetism, because the latter tends to align spins in the same direction, which is unfavourable for the singlet Cooper pairing\cite{DeGennes}. On the other hand, in superfluid Fermi gases, `spins' in Cooper pairs are actually {\it pseudospins} describing two atomic hyperfine states. Thus, at present, there is no object corresponding to the magnetic impurity in superconductivity. Thus, if we can introduce a {\it magnetic object} to a cold Fermi gas, together with the tunable interaction and optical lattice, this system would become a more useful quantum simulator to study metallic superconductivity.
\par
\begin{figure}[t]   
\begin{center}
\includegraphics[keepaspectratio,scale=0.8]{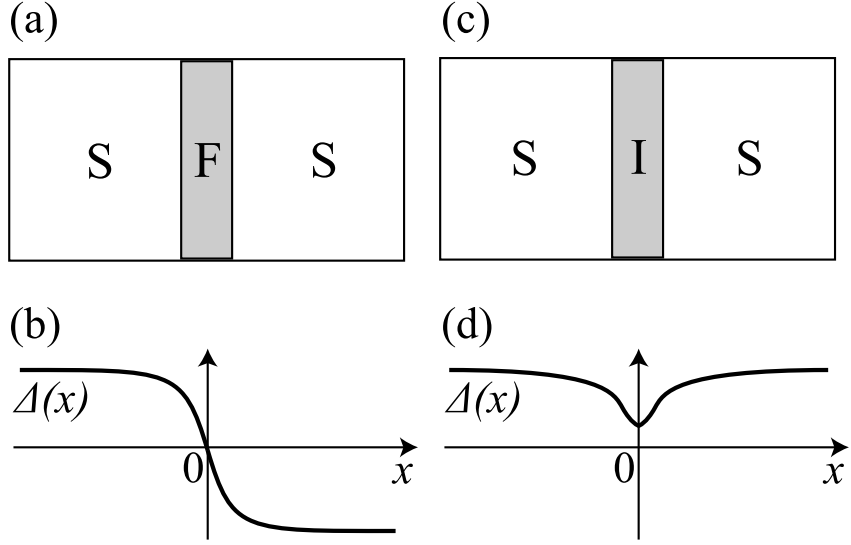}
\caption{(a) Schematic SFS-junction. `S' and `F' mean the superfluid (superconducting) region and ferromagnet, respectively. (b) Spatial variation of superfluid order parameter $\Delta(x)$ in the $\pi$-phase. (c) SIS-junction, where `I' means the insulating barrier. (d) Spatial variation of $\Delta(x)$ in the SIS-junction (0-phase).}
\label{fig.1}
\end{center}    
\end{figure}

In this paper, we theoretically discuss an idea which enables us to study magnetic effects on superfluid Fermi gases. Among various magnetic effects discussed in metallic superconductivity, in this paper, we focus on a superfluid/ferromagnet/superfluid (SFS)-junction. In superconductivity, the SFS-junction in Fig.\ref{fig.1}(a) is well known to modulate the phase of superconducting order parameter. Namely, it induces the $\pi$-phase\cite{Bulaevskii,Buzdin,Kanegae,Ryazanov,SFSexp} shown in Fig.\ref{fig.1}(b), where the order parameter changes its sign across the SFS-junction. Such a phase modulation does not occur in the superconductor/insulator/superconductor (SIS)-junction in Fig.\ref{fig.1}(c), where what we call the 0-phase shown in Fig.\ref{fig.1}(d) is realized.
\par
The origin of $\pi$-phase in the superconducting SFS-junction is deeply related to the FFLO state. When the order parameter $\Delta$ penetrates into the ferromagnetic junction, a molecular field in the ferromagnet induces an FFLO-like spatial oscillation of $\Delta$. Inside the junction, when the sign of oscillating order parameter at the right edge of the junction is opposite to the sign of the order parameter at the left edge, the $\pi$-phase is realized. On the other hand, when the order parameter has the same sign at both the edges, the $0$-phase is obtained. Indeed, a transition between $0$-phase and $\pi$-phase (0-$\pi$ transition) has been observed in Nb/Cu$_{0.47}$Ni$_{0.53}$/Nb-junction, when one varies the thickness of ferromagnetic Cu$_{0.47}$Ni$_{0.53}$-layer\cite{SFSexp}.
\par

\begin{figure}[t]   
\begin{center}
\includegraphics[keepaspectratio, scale=0.8]{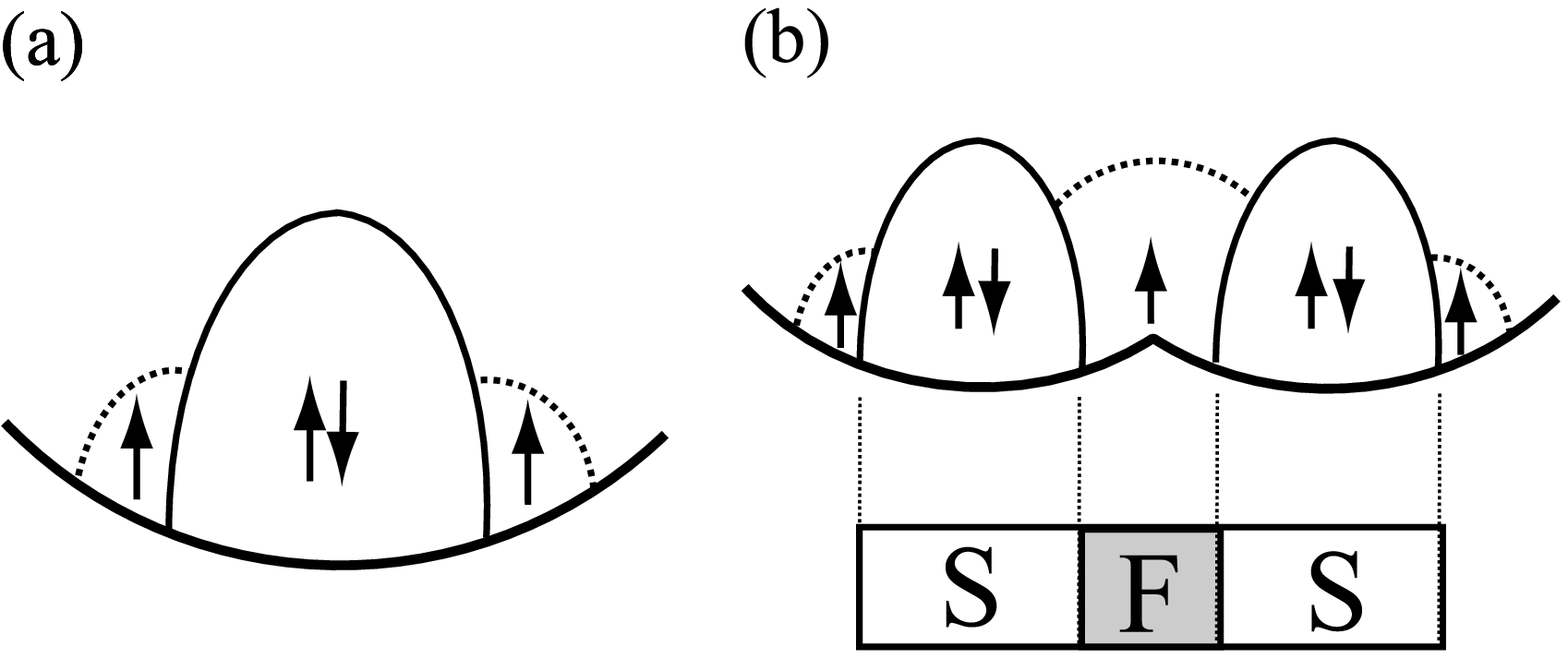}
\caption{(a) Schematic density profile of a trapped superfluid Fermi gas with population imbalance ($N_\uparrow>N_\downarrow$). Excess $\uparrow$-spin atoms are localized at the edges of superfluid region ($\uparrow\downarrow$). (b) Density profile expected in a double well potential. Excess atoms can be also localized around the central potential barrier. When we regard the central spin-polarized region as a ferromagnet, the system can be viewed as an SFS-junction.}
\label{fig.2}
\end{center}    
\end{figure}

\par
We explain the outline of our idea. In a superfluid Fermi gas with population imbalance, the phase separation has been observed\cite{MIT3,RICE}, as schematically shown in Fig.\ref{fig.2}(a). Then, noting that excess atoms tend to occupy the spatial region where potential energy is large\cite{MIT3,RICE,Silva,Chevy,Pieri,Yi,Silva2,Haque,Tezuka}, we expect that, in a double well potential, excess atoms are also localized around the central potential barrier, as shown in Fig.\ref{fig.2}(b). Regarding this central region as a {\it ferromagnet of pseudospin-$\uparrow$}, Fig.\ref{fig.2}(b) looks like an SFS-junction. Although this ferromagnet is actually a gas of atoms in the hyperfine state described by pseudospin-$\uparrow$, we will find that it indeed induces the $\pi$-phase as in the case of real ferromagnet.  
\par
This paper is organized as follows: In Sec.II, we explain our formulation. Treating an attractive Hubbard model within the mean field theory at $T=0$, we numerically determine the superfluid order parameter, as well as the particle density, around a potential barrier in the presence of population imbalance. We compare the energy of $\pi$-phase with that of $0$-phase to examine the stability of $\pi$-phase. In Sec. III, we examine the possibility of SFS-junction and $\pi$-phase in a simple one-dimensional model in the absence of a trap. Effects of a harmonic trap, as well as dimensionality, will be discussed in Sec.IV.
\par


\section{Formulation}
\par
We consider a two-component Fermi gas with population imbalance. To examine the possibility of SFS-junction and $\pi$-phase in a simple manner, we employ the Hubbard model described by the Hamiltonian,
\begin{equation}
H = -t \sum_{\ev{i,j},\sigma}
\left[
\hat{c}_{i,\sigma}^\dagger \hat{c}_{j,\sigma} + {\rm h.c.}
\right] 
-U 
\sum_{i} \hat{n}_{i,\uparrow } \hat{n}_{i,\downarrow}
+\sum_{i,\sigma}
\left[V_i-\mu_\sigma
\right]\hat{n}_{i,\sigma }.
\label{H}
\end{equation}
Here, $\hat{c}^\dagger_{i,\sigma}$ is the creation operator of a Fermi atom at the $i$-th site, where the pseudospin $\sigma=\uparrow,\downarrow$ describes two atomic hyperfine states. $t$ is a nearest-neighbour hopping, and the summation $\ev{i,j}$ is taken over the nearest-neighbour pairs. $-U (<0)$ is an on-site pairing interaction, and $\hat{n}_{i,\sigma}=\hat{c}_{i,\sigma}^\dagger \hat{c}_{i,\sigma}$ is the number operator at the $i$-th site. Since we consider a polarized Fermi gas, the chemical potential $\mu_\sigma$ depends on pseudospin $\sigma$. $V_i$ is a (non-magnetic) potential, consisting of a potential barrier $V_i^b$ and a harmonic trap $V_i^t$. Their detailed expressions are given later.
\par
Although a real superfluid Fermi gas is a three-dimensional continuum system, we consider a one-dimensional chain and two-dimensional square lattice in this paper. In this regard, we emphasize that the assumed low dimensionality and lattice structure are not essential for the realization of SFS-junction and $\pi$-phase discussed in this paper. However, to suppress band effects by the lattice, we consider the case of low particle density $\langle \hat{n}_{i,\sigma}\rangle\lesssim 0.3$.
\par
We treat Eq.(\ref{H}) within the mean field theory at $T=0$.  Under this approximation, we have
\begin{eqnarray}
H_\text{MF} 
&=& 
-t \sum_{\langle i,j \rangle,\sigma} 
\left[
\hat{c}^\dagger_{i,\sigma} \hat{c}_{j,\sigma}+ {\rm h.c.} 
\right] 
-\sum_i
\left[
\Delta_i\hat{c}_{i,\up}^\dagger \hat{c}_{i,\dwn}^\dagger 
+ \Delta_i^*\hat{c}_{i,\dwn} \hat{c}_{i,\up}
\right]
\nonumber
\\
&+&\sum_{i,\sigma} 
\left[
V_i-\mu_\sigma-U \ev{\hat{n}_{i,-\s}} 
\right]
\hat{n}_{i,\s}
+ \sum_{i} 
\left[
\frac{|\op_i|^2}{U} + U \ev{\hat{n}_{i,\up}} 
\ev{\hat{n}_{i,\dwn}} 
\right].
\label{MFH}
\end{eqnarray}
Here, $\Delta_i=U\langle \hat{c}_{i,\downarrow}\hat{c}_{i,\uparrow}\rangle$ is the superfluid order parameter (which is taken to be real in the following calculations). Equation (\ref{MFH}) can be rewritten as
\begin{equation}
H_\text{MF} = 
\hat{\Psi}^\dagger \tilde{H} \hat{\Psi}
+ \sum_{i} 
\left[
V_i - \cp_\dwn -U \ev{\hat{n}_{i,\up}} 
+ \frac{\op_i^2}{U} 
+ U \ev{\hat{n}_{i,\up}} \ev{\hat{n}_{i,\dwn}} 
\right].
\label{NambuRep}
\end{equation}
Here, $\hat{\Psi}^\dagger =[\hat{c}_{1,\up}^\dagger~ \hat{c}_{2,\up}^\dagger~\cdots~ \hat{c}_{M,\up}^\dagger~\hat{c}_{1,\dwn}~ \hat{c}_{2,\dwn}~\cdots~ \hat{c}_{M,\dwn}]$, where $M$ is the total number of lattice sites. The matrix elements of $\tilde{H}$ are chosen so that Eq.(\ref{NambuRep}) can reproduce Eq.(\ref{MFH}). As usual, Eq.(\ref{NambuRep}) can be conveniently diagonalized by the Bogoliubov transformation,
\begin{eqnarray}
\left(
\begin{array}{c}
{\hat c}_{1,\uparrow}\\
{\hat c}_{2,\uparrow}\\
\vdots\\
{\hat c}_{M,\uparrow}\\
{\hat c}_{1,\downarrow}^\dagger\\
{\hat c}_{2,\downarrow}^\dagger\\
\vdots\\
{\hat c}_{M,\downarrow}^\dagger\\
\end{array}
\right)
=
{\hat W}
\left(
\begin{array}{c}
{\hat \alpha}_{1,\uparrow}\\
{\hat \alpha}_{2,\uparrow}\\
\vdots\\
{\hat \alpha}_{M,\uparrow}\\
{\hat \alpha}_{1,\downarrow}^\dagger\\
{\hat \alpha}_{2,\downarrow}^\dagger\\
\vdots\\
{\hat \alpha}_{M,\downarrow}^\dagger\\
\end{array}
\right),
\label{eq.BB}
\end{eqnarray}
where ${\hat W}$ is a $2M\times 2M$-orthogonal matrix. The diagonalized Hamiltonian has the form
\begin{equation}
H_\text{MF}
=\sum_{j,\sigma}E_{j,\s} \hat{\alpha}_{j,\s}^\dagger \hat{\alpha}_{j,\s} 
+ \sum_{i=1}^M 
\left[
V_i-\cpdwn - U\ev{\hat{n}_{i,\up}} + \frac{\op_i^2}{U}
+ U \ev{\hat{n}_{i,\up}} \ev{\hat{n}_{i,\dwn}} 
 -E_{i,\downarrow}
\right],
\label{DiagonalH}
\end{equation}
where $E_{j,\sigma}$ is the Bogoliubov excitation spectrum. The superfluid order parameter $\Delta_i=U\langle \hat{c}_{i,\downarrow}\hat{c}_{i,\uparrow}\rangle$ and particle density $\langle \hat{n}_{i,\sigma}\rangle$ in Eq.(\ref{DiagonalH}) are given by, respectively,
\begin{equation}
\Delta_i = U \sum_{j=1}^M
\Bigl[
W_{i,j} W_{M+i,j} \theta( -E_{j,\up} )+ W_{i,M+j} W_{M+i,M+j} 
\theta(E_{j,\downarrow})
\Bigr],
\label{GAP}
\end{equation}
\begin{eqnarray}
\begin{array}{l}
\displaystyle
\ev{\hat{n}_{i,\up}} = 
\sum_{j=1}^M 
\Bigl[
W_{i,j}^2 \theta( -E_{j,\up} )+ W_{i,M+j}^2 \theta(E_{j,\downarrow})
\Bigr], 
\\
\displaystyle
\ev{\hat{n}_{i,\dwn}} =
\sum_{j=1}^M 
\Bigl[
W_{M+i,j}^2 \theta( E_{j,\up} )+ W_{M+i,M+j}^2 \theta(-E_{j,\downarrow}) 
\Bigr].
\end{array}
\label{DENSITY}
\end{eqnarray}
In the presence of population imbalance, since $E_{j,\sigma}$ may be negative, the step functions appear in Eqs.(\ref{GAP}) and (\ref{DENSITY}). In the unpolarized case, one can safely take $E_{j,\uparrow}=E_{j,\downarrow}>0$\cite{DeGennes}.
\par
We numerically diagonalize ${\tilde H}$, and self-consistently determine $\Delta_i$, $\ev{\hat{n}_{i,\sigma}}$, as well as $\mu_\sigma$, for a given parameter set $(U,N_\uparrow,N_\downarrow,V_i)$, where $N_\sigma=\sum_i \ev{\hat{n}_{i,\sigma}}$ is the total number of $\sigma$-spin atoms. Once the self-consistent $\pi$-phase solution is obtained, we energetically compare this state with the 0-phase solution. In this procedure, since these two self-consistent solutions usually have different values of chemical potentials $\mu_\sigma$, it is convenient to use the free energy $E_G$ rather than the thermodynamic potential $\Omega$ ($=\langle H_{\rm MF}\rangle$ at $T=0$). The former is immediately obtained from the latter by the Legendre transformation, as $E_G=\Omega+\sum_\sigma\mu_\sigma N_\sigma$. At $T=0$, we have
\begin{eqnarray}
E_G
&=&
\sum_{j,\sigma}E_{j,\sigma} \theta (-E_{j,\sigma})
+ \sum_{\s} \cp_\s N_\s
\nonumber
\\
&+& \sum_{i} 
\left[ V_i - \mu_\downarrow - U \langle \hat{n}_{i,\uparrow } \rangle
+ {\Delta_i^2 \over U}  
+ U \langle \hat{n}_{i,\uparrow } \rangle 
\langle \hat{n}_{i,\downarrow} \rangle 
 - E_{i,\downarrow}
\right].
\label{Legendre}
\end{eqnarray}
We will use the energy difference $\Delta E_G\equiv E_G^{\pi{\rm -phase}}-E_G^{\rm 0-phase}$ between the $\pi$-phase and $0$-phase to examine the stability of $\pi$-phase.
\par

\begin{figure}[t]
\begin{center}
\includegraphics[keepaspectratio, scale=0.8]{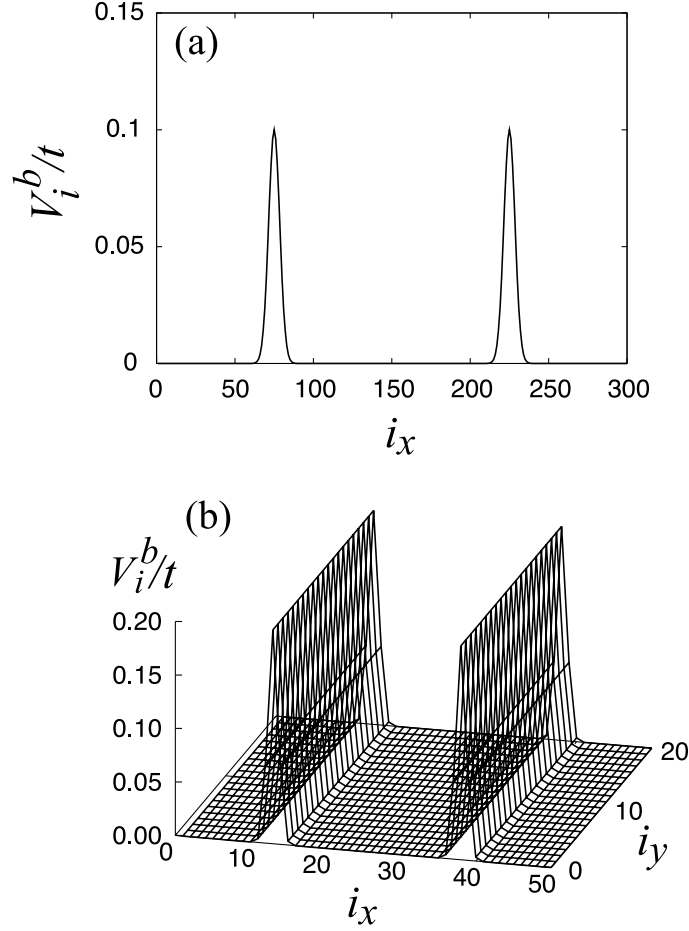}    
\caption{Potential barrier $V_i^b$ used in this paper. (a) One-dimensional model. (b) Two-dimensional model. $i_x$ and $i_y$ represent the site number in the $x$- and $y$-direction, respectively.}
\label{fig.3}
\end{center}
\end{figure}

Here, we present the detailed expression for the potential $V_i=V_i^t+V_i^b$. In Sec.III, we ignore effects of a trap $V_i^t$ for simplicity, and only take into account the potential barrier $V_i^b$ in the one-dimensional system. However, considering the model with only one potential barrier under the periodic boundary condition, we cannot obtain the $\pi$-phase solution, because it does not satisfy the periodic boundary condition imposed at the edges of the system. To avoid this problem, in this paper, we consider the system with {\it two} potential barriers shown in Fig.\ref{fig.3}(a). Namely, we set
\begin{eqnarray}
V^b_i=
\left\{
\begin{array}{ll}
V_0^b e^{-(i_x-L_x/4)^2/\ell^2}&
~~~~1   \leq i_x < L_x/2, \\
V_0^b e^{-(i_x-3L_x/4)^2/\ell^2}&
~~~~L_x/2 \leq i_x \leq L_x,
\end{array}
\right.
\label{1DPot}
\end{eqnarray} 
where $L_x$ is the number of lattice sites in the $x$-direction, and $i_x$ is the site number. (The lattice constant is taken to be unity.) $V_0^b$ and $\ell$ describe the height and width of the potential barrier, respectively. We briefly note that the assumed Gaussian spatial dependence in Eq.(\ref{1DPot}) is not crucial for the formation of SFS-junction and $\pi$-phase. Using Eq.(\ref{1DPot}), we can conveniently obtain both the $0$-phase and $\pi$-phase solutions under the same periodic boundary condition. We also use this prescription in considering the two-dimensional model in Fig.\ref{fig.3}(b).
\par
In Sec. IV, we include the harmonic trap $V_i^t$. In the one-dimensional chain, we set $V_i$ as
\begin{equation} 
V_i = V_0^t \left(i_x-{L_x \over 2}\right)^2+V_0^b e^{-(i_x-L_x/2)^2/\ell^2},
\label{trap1D}
\end{equation}
where $V_0^t$ describes the strength of harmonic trap. The potential barrier $V_i^b$ is centered at $i_x=L_x/2$. In the case of two-dimensional cigar trap, we take
\begin{equation} 
V_i= 
\left[
V_{0x}^t \left(i_x-{L_x \over 2}\right)^2 
+ V_{0y}^t \left(i_y-{L_y \over 2}\right)^2
\right]
+V_0^b e^{-(i_x-L_x/2)^2/\ell^2},
\label{trap2D}
\end{equation}
where the cigar trap is described by $(V_{0x}^t,V_{0y}^t)$, and $L_y$ is the number of lattice sites in the $y$-direction. In Eq.(\ref{trap2D}), the potential barrier $V_i^b$ is centered at $i_x=L_x/2$ and parallel to the $y$-axis.
\par

\begin{figure}[t]
\begin{center}
\includegraphics[keepaspectratio, scale=0.8]{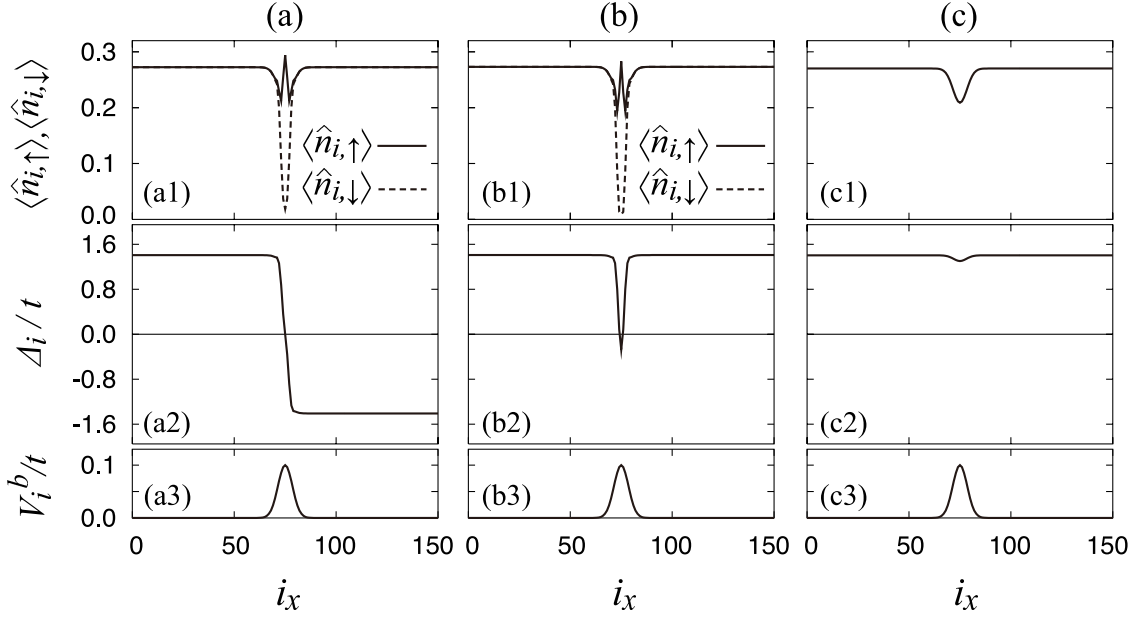}
\caption{Self-consistent $\pi$-phase solution (a) and $0$-phase solution (b) in the one-dimensional model in Fig.\ref{fig.3}(a). From the comparison of their energies, we find that the $\pi$-phase is more stable than the $0$-phase. We also show the case of unpolarized Fermi superfluid in (c). The particle density $\langle n_{i,\sigma}\rangle$ and superfluid order parameter $\Delta_i$ are shown in the upper and middle panels, respectively. For convenience, the potential barrier $V_i^b$ is also shown in the lower panels. The total number of atoms equals 160, and we take $N_\up=81$, $N_\dwn=79$ for (a) and (b), and $N_\up=N_\dwn=80$ for (c). The other parameters are taken as $U/t=4$\cite{note}, $L_x=300$, $V_0^b/t=0.1$ and $\ell=5$. The same system size and parameters for the barrier are also used in Figs.\ref{fig.5}-\ref{fig.9}. Since results are always symmetric with respect to $i_x=L_x/2$, we only show the region $1\le i_x\le L_x/2$ here.
}
\label{fig.4}
\end{center}
\end{figure}
\par
\begin{figure}[t]   
\begin{center}
\includegraphics[keepaspectratio, scale=0.8]{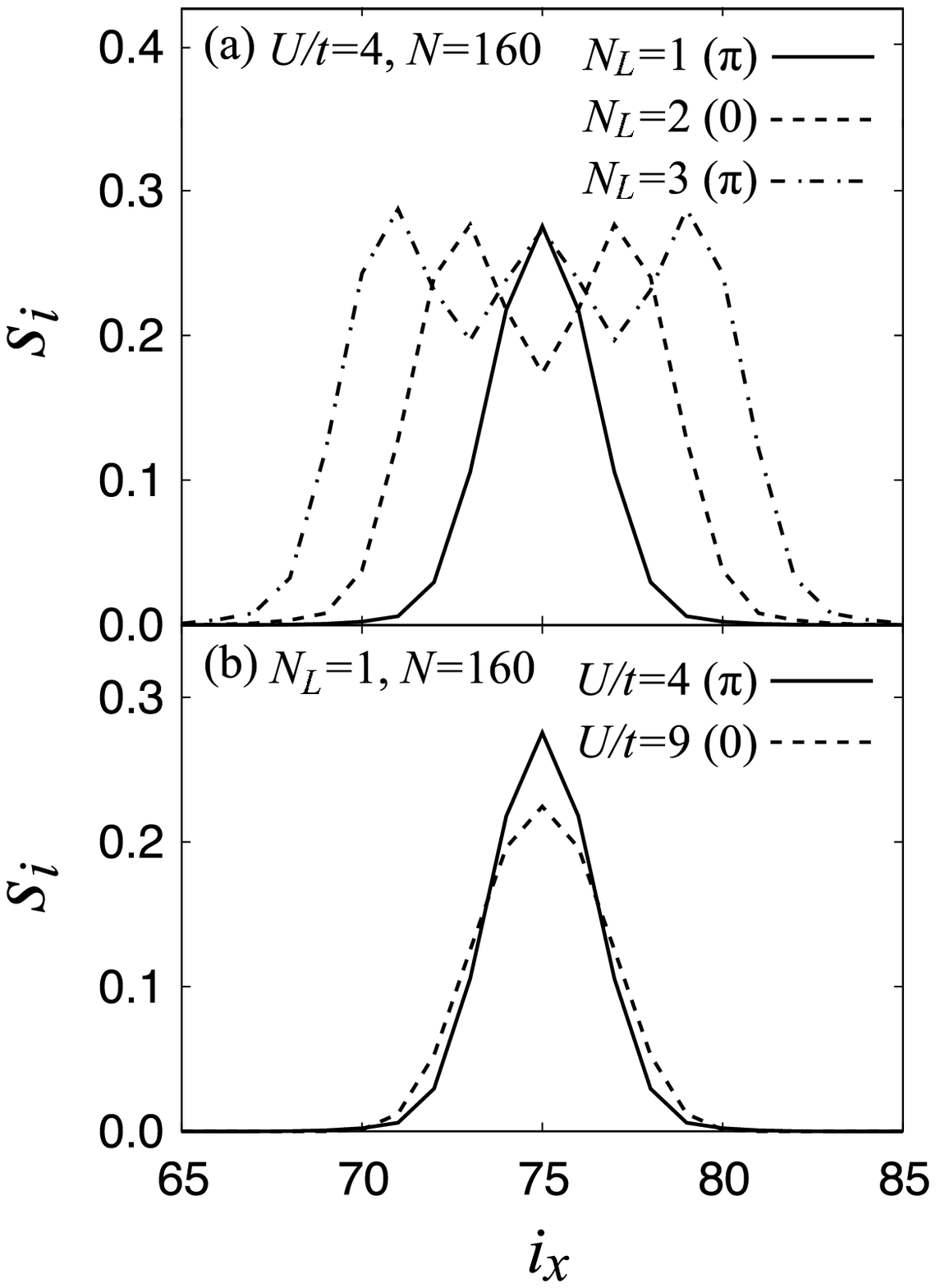}    
\caption{Spatial variation of polarization $s_i=\langle n_{i,\uparrow}\rangle -\langle n_{i,\downarrow}\rangle$ around the barrier. (a) shows effects of the number of localized excess atoms $N_L$. (b) shows effects of the pairing interaction $U$. $``\pi"$ $(``0")$ in the legend means that the $\pi$-phase (0-phase) is the the most stable state in each case. 
}
\label{fig.5}
\end{center}    
\end{figure}

\section{Formation of SFS-junction and $\pi$-phase}
\par
Figure \ref{fig.4} shows the self-consistent $\pi$-phase solution (a) and $0$-phase solution (b) in the one-dimensional model in Fig.\ref{fig.3}(a). As expected, the polarization $s_i\equiv \langle n_{i,\uparrow}\rangle-\langle n_{i.\downarrow}\rangle$ is finite around the barrier, which may be viewed as a ferromagnetic junction. In Fig.\ref{fig.4} (where $N_\uparrow-N_\downarrow=2$), since there are two potential barriers in the model, the number of excess atoms localized around each barrier ($\equiv N_L$) equals one. Strictly speaking, there is no clear boundary between the superfluid region and spin-polarized region because of the continuous variations of $\Delta_i$ and $s_i$. In this paper, we simply call the region where $s_i$ is large to some extent the ferromagnetic junction. For example, in the case of Fig.\ref{fig.4}, from the spatial variation of $s_i$ shown in Fig.\ref{fig.5}, we roughly regard the region $72\lesssim i_x\lesssim 78$ as the ferromagnetic junction. 
\par
\begin{figure}[t]   
\begin{center}
\includegraphics[keepaspectratio, scale=0.8]{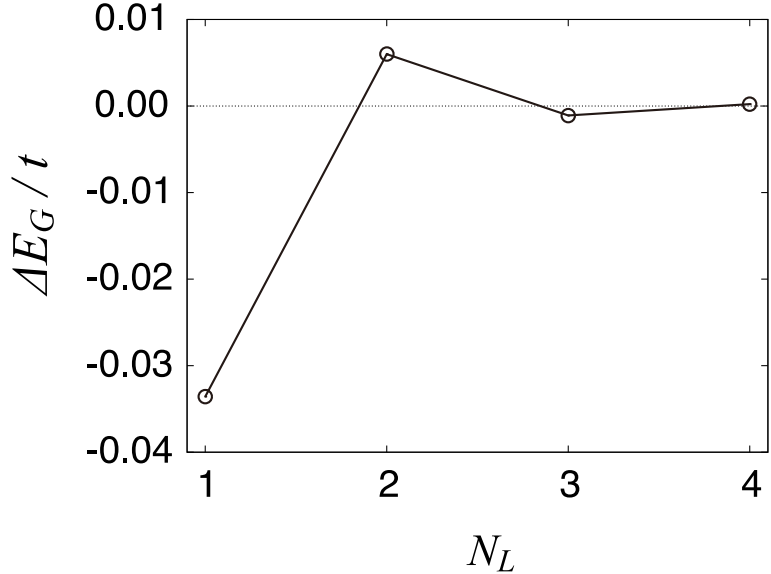}    
\caption{Energy difference $\Delta E_G=E_G^{\pi{\rm -phase}}-E_G^{\rm 0-phase}$ between the $\pi$-phase and 0-phase, as a function of the number of localized excess atoms per one potential barrier. Since there are two barriers in Fig.\ref{fig.3}(a), $N_L$ equals $[N_\uparrow-N_\downarrow]/2$. We take $U/t=4$ and $N=160$.
}
\label{fig.6}
\end{center}    
\end{figure}

In the $\pi$-phase solution (Fig.\ref{fig.4}(a)), the order parameter $\Delta_i$ changes its sign across the ferromagnetic junction, which is in contrast to the $0$-phase case (Fig.\ref{fig.4}(b)). From the comparison of their energies, the $\pi$-phase is found to be more stable than the $0$-phase. (Note that $\Delta E_G<0$ at $N_L=1$ in Fig.\ref{fig.6}.)  This confirms that the spin-polarized region really works as a ferromagnetic junction. 
\par
\par
We note that the $\pi$-phase is not obtained in the absence of population imbalance. As an example, Fig.\ref{fig.4}(c) shows the case of $N_\uparrow=N_\downarrow=80$, where the order parameter (panel (c2)) is found to be not affected by the barrier, except for a slight decrease around it due to the slight decrease of particle density shown in Fig.\ref{fig.4}(c1)\cite{note2}. 
\par
The reason why excess $\uparrow$-spin atoms are localized around the barrier $V_i^b$ is understood as follows. When the phase separation occurs in a polarized Fermi superfluid, the superfluid order parameter is destroyed in the spatial region where excess atoms are localized. Then, to make the loss of condensation energy by this pair-breaking effect as small as possible, the localization should occur around the potential barrier, where the order parameter is already (slightly) suppressed in the unpolarized case, as shown in Fig.\ref{fig.4}(c2).
\par

\begin{figure}[t]   
\begin{center}
\includegraphics[keepaspectratio, scale=0.8]{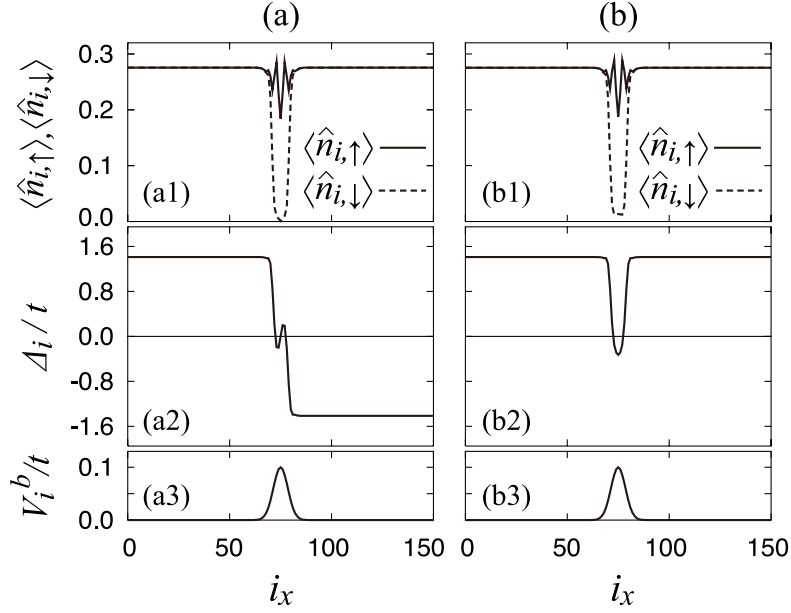}    
\caption{(a) $\pi$-phase and (b) $0$-phase solutions when $N_\up=82$, $N_\dwn=78$, and $U/t=4$. In this case, two excess atoms are localized around the potential barrier, and the $0$-phase is more stable than the $\pi$-phase.}
\label{fig.7}
\end{center}
\end{figure}

\begin{figure}[t]   
\begin{center}
\includegraphics[keepaspectratio, scale=0.8]{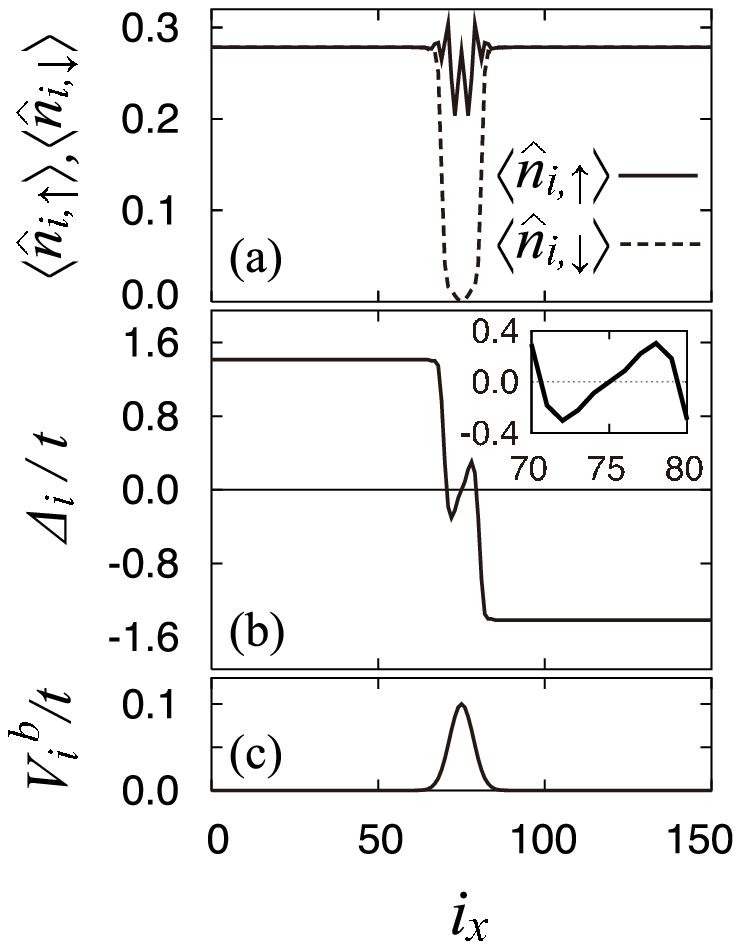}    
\caption{Stable $\pi$-phase solution when $N_L=3$. We take $N_\up=83$, $N_\dwn=77$, and $U/t=4$. The inset shows the magnified spatial variation of $\Delta_i$ in the ferromagnetic junction ($70\lesssim i_x \lesssim 80$).
} 
\label{fig.8}
\end{center}    
\end{figure}
\par
As explained in the introduction, the $\pi$-phase originates from the FFLO-type spatial oscillation of order parameter in the ferromagnetic junction. Because of this mechanism, the 0-$\pi$ transition has been observed in superconducting Nb/Cu$_{0.47}$Ni$_{0.53}$/Nb-junction\cite{SFSexp}, as one varies the thickness of ferromagnetic Cu$_{0.47}$Ni$_{0.53}$-layer. In the present superfluid Fermi gas, since the thickness of ferromagnetic junction depends on $N_L$ (See Fig.\ref{fig.5}(a).), we can also expect the same phenomenon. Indeed, Fig.\ref{fig.6} shows that the 0-phase becomes more stable than the $\pi$-phase when two excess $\uparrow$-spin atoms are localized around the potential barrier ($N_L=2$). We show the self-consistent solutions in this case in Fig.\ref{fig.7}. In this case, since the 0-phase is the most stable state, to make the $\pi$-phase solution, one needs to forcibly {\it bend} the order parameter in the ferromagnetic junction, which leads to the fine structure around the barrier seen in Fig.\ref{fig.7}(a2).
\par
Figure \ref{fig.6} shows that the $\pi$-phase again becomes more stable than the 0-phase when $N_L=3$. In this case, we see in Fig.\ref{fig.8}(b) an FFLO-like oscillation with the wavelength $\lambda\simeq 9$ in the ferromagnetic junction ($70\lesssim i_x\lesssim 80$). In the ordinary FFLO state, the wavelength of the order parameter oscillation is given by\cite{Takada},
\begin{equation}
\lambda\simeq 2\pi{v_{\rm F} \over h},
\label{eq.FFLO}
\end{equation}
where $v_{\rm F}$ and $h$ are the Fermi velocity and an external magnetic field, respectively (where $h$ involves the Bohr magneton). In the present model Hamiltonian in Eq.(\ref{MFH}), $h$ in the ferromagnetic junction is evaluated as, 
\begin{equation}
h\simeq {1 \over 2}
\left[
\mu_\uparrow-\mu_\downarrow
-U
\overline{s_i}
\right],
\label{eq.FFLO2}
\end{equation} 
where $\overline{s_i}$ is the polarization averaged over the junction. Substituting numerical values into Eqs.(\ref{eq.FFLO}) and (\ref{eq.FFLO2})\cite{note4}, we obtain $\lambda\simeq 9.1$, which is consistent with the value ($\lambda\simeq 9$) evaluated from the inset in Fig.\ref{fig.8}(b).
\par
When the ferromagnetic junction becomes very thick, the Josephson coupling between the two superfluid regions becomes very weak. Since the energy difference $\Delta E_G$ between the $\pi$-phase and $0$-phase purely comes from the Josephson coupling, $|\Delta E_G|$ in Fig.\ref{fig.6} becomes small with increasing $N_L$.
\par

\begin{figure}[t]   
\begin{center}
\includegraphics[keepaspectratio, scale=0.8]{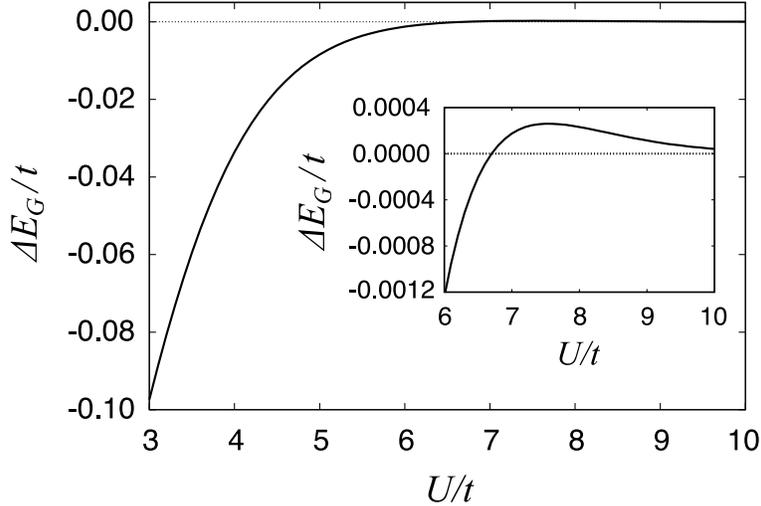}    
\caption{Energy difference $\Delta E_G$ between $\pi$-phase and 0-phase as a function of $U$. We take $N_\uparrow = 81$ and $N_\downarrow=79$. The inset shows $\Delta E_G$ magnified around the 0-$\pi$ transition.} 
\label{fig.9}
\end{center}    
\end{figure}

\par
While the thickness of the ferromagnetic junction depends on $N_L$, Fig.\ref{fig.5}(b) shows that it is not so sensitive to the pairing interaction $U$. However, even in this case, Fig.\ref{fig.9} shows that the 0-$\pi$ transition occurs when $U$ becomes large. This is because, when the BCS excitation gap becomes large due to strong pairing interaction, the {\it effective magnetic field} given by Eq.(\ref{eq.FFLO2}) also becomes large to realize a finite population imbalance\cite{noteH}. Thus, the FFLO wavelength $\lambda$ in Eq.(\ref{eq.FFLO}) becomes short, leading to the transition from the $\pi$-phase to the 0-phase. We briefly note that, while the 0-$\pi$ transition occurs by tuning $N_L$ and $U$, it does not occur when the potential shape is varied by tuning $V_0^b$ and $\ell$, at least within our numerical calculations (although we do not show the results here). This means that the dominant role of potential barrier is a pinning center. Once excess atoms are pinned at the potential barrier, the magnetic effect is dominated by the number of localized atoms $N_L$, as well as the pairing interaction $U$.
\par

\begin{figure}[t]   
\begin{center}
\includegraphics[keepaspectratio, scale=0.6]{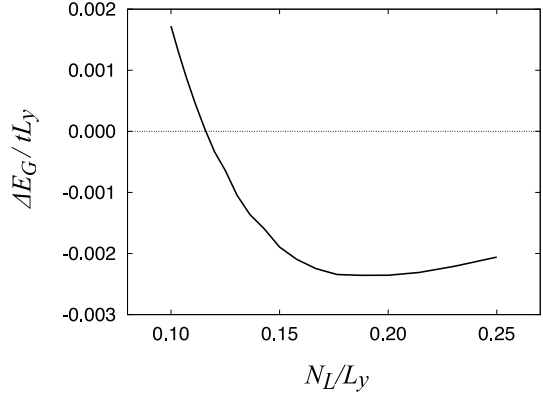}    
\caption{Energy difference $\Delta E_G$ as a function of $N_L/L_y$, in the two-dimensional model. The $x$-axis means the number of localized excess atoms around the barrier per unit length in the $y$-direction. We set $U/t=7$, $V_0^t/t=0.2$, $\ell=1$, and $L_x=50$, which are also used in Fig.\ref{fig.11}. For the number of atoms, we take $N_\sigma=(N+2\sigma N_L)/2$ and $N=6L_y$. The value of $L_y$ ($=17\sim 48$) is chosen so that $L_y/N_L$ can be an integer. 
}
\label{fig.10}
\end{center}    
\end{figure}

\begin{figure}[t]   
\begin{center}
\includegraphics[keepaspectratio, scale=0.6]{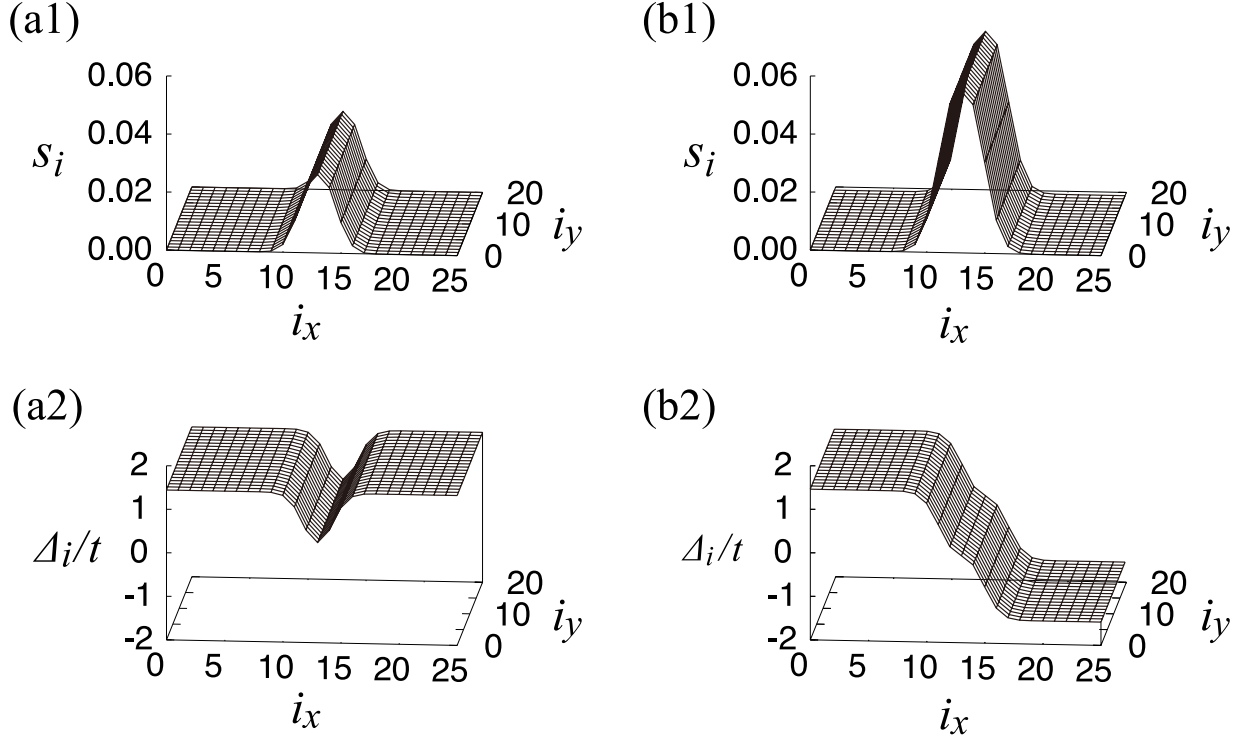}
\caption{(a) Stable 0-phase solution when $N_L/L_y=0.1$ ($L_y=30$). (b) Stable $\pi$-phase solution when $N_L/L_y= 0.25$. ($L_y=20$). The upper and lower panels show the spatial variations of $s_i$ and $\Delta_i$, respectively. We only show the spatial region $1\le i_x\le 25$ and $1\le i_y\le 20$ in this figure.
}
\label{fig.11}
\end{center}    
\end{figure}

\section{Effects of dimensionality and trap potential}
\par
In the previous section, we considered the one-dimensional model in the absence of a trap, to simply examine the possibility of SFS-junction and $\pi$-phase in a superfluid Fermi gas. In this section, we extend the previous study to include higher dimension and harmonic trap.
\par
\subsection{SFS-junction and $\pi$-phase in two dimension}
\par
In the one-dimensional model discussed in Sec. III, although the SFS-junction is obtained, the stability of $\pi$-phase is sensitive to the number $N_L$ of excess atoms localized around the potential barrier. This means that we have to adjust $N_L$ within the accuracy of $O(1)$ in order to realize the $\pi$-phase, which is actually difficult experimentally. However, in a higher dimensional system, since excess atoms can align in the $y$- and $z$-direction without remarkably increasing the thickness of ferromagnetic junction, the $\pi$-phase is expected to be more accessible.
\par
To see this, we examine the two-dimensional model shown in Fig.\ref{fig.3}(b). As expected, we obtain the finite region of stable $\pi$-phase in Fig.\ref{fig.10} ($N_L/L_y\gesim 0.12$). Comparing Figs.\ref{fig.11}(a1) with (b1), we find that, while the increase of localized atoms results in the enhancement of polarization $s_i$ at the center of the barrier, the thickness of the ferromagnetic region is almost unchanged. However, since the junction would eventually become thick when one increases the number of excess atoms very much, the upper bound of the $\pi$-phase region in Fig.\ref{fig.10} would exist at a large value of $N_L/L_y ~(>0.25)$.
\par

\begin{figure}[t]   
\begin{center}
\includegraphics[keepaspectratio, scale=0.9]{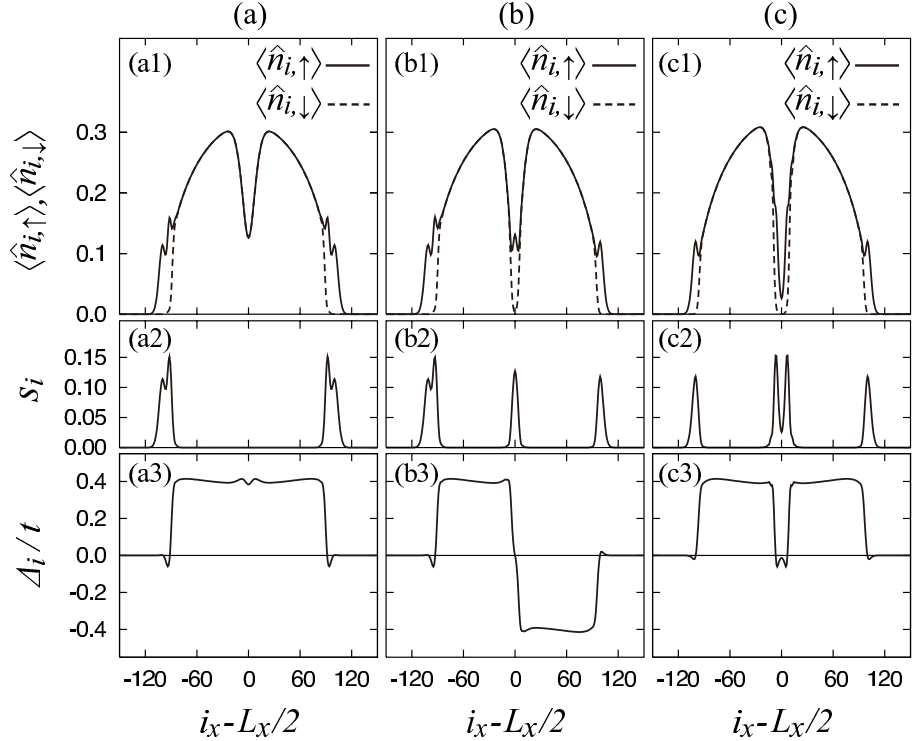}    
\caption{One-dimensional polarized superfluid Fermi gas trapped in a harmonic potential. The height of the potential barrier $V_0^b$ is taken to be (a) $0.4t$, (b) $0.5t$, and (c) $0.6t$. We take $N_\up=47$, $N_\dwn=43$, $U/t=2$, $V_0^t=0.00005t$, and $\ell=12$. The density profile $\langle n_{i,\sigma}\rangle$, polarization $s_i$, and $\Delta_i$ are shown in the upper, middle, and lower panels, respectively. The potential barrier is centered at $i_x=L_x/2=150$.
}
\label{fig.12}
\end{center}
\end{figure}

\begin{figure}[t]   
\begin{center}
\includegraphics[keepaspectratio, scale=0.9]{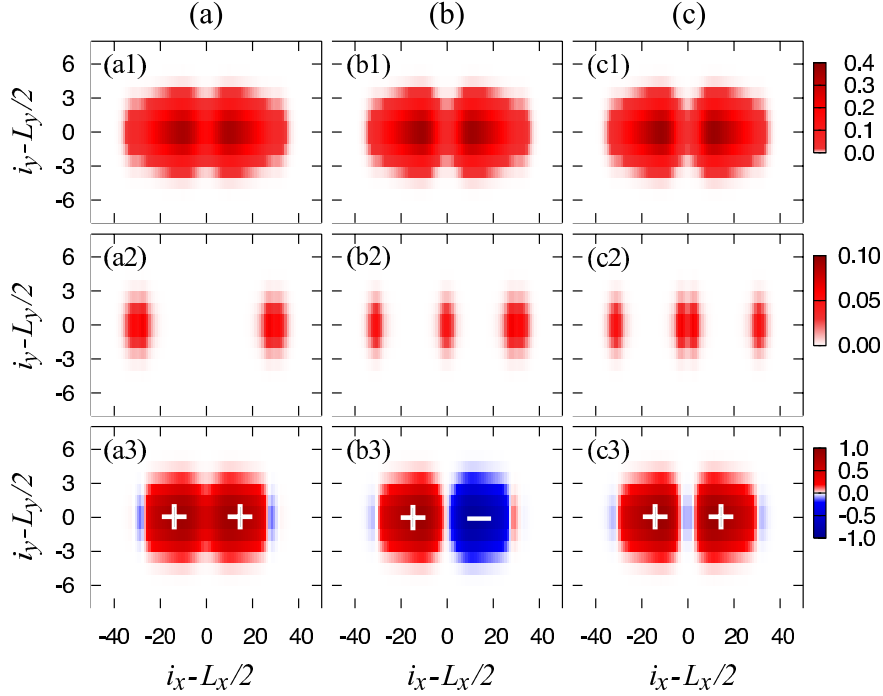} 
\caption{(Color online) Two-dimensional polarized superfluid Fermi gas in a two-dimensional cigar trap. The central barrier height $V_0^b$ is given by (a) $0.6t$, (b) $1.0t$, and (c) $1.2t$. For other parameters, we take $N_\up=32$, $N_\dwn=28$, $U/t=4$, $V_{0x}^t/t=0.001$, $V_{0y}^t/t=0.1$, and $\ell=6$. The upper panels show the total density profile, and the middle panels show the polarization. The order parameter is shown in the lower panels. In panels (b), the number of localized excess atoms around the central barrier equals one, and $\pi$-phase is obtained as the most stable state. In panels (c), the number of excess atoms around the barrier equals two, where the 0-phase is realized. For clarity, we explicitly write the sign of the order parameter ("+" and "-") in the lower panels.}
\label{fig.13}
\end{center}    
\end{figure}

\begin{figure}[t]   
\begin{center}
\includegraphics[keepaspectratio, scale=0.5]{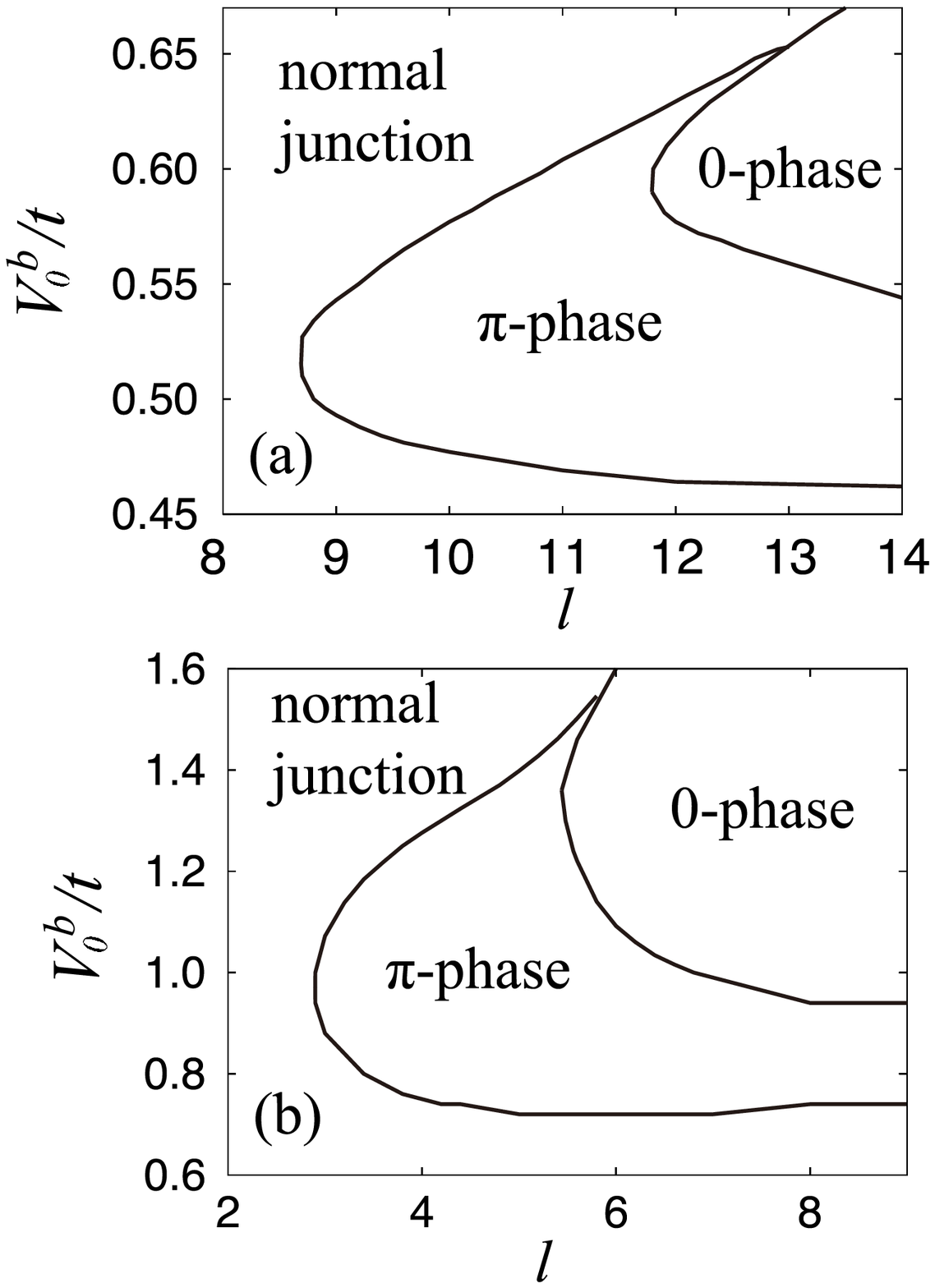}    
\caption{Phase diagram of trapped polarized Fermi superfluid in terms of potential height $V_0^b$ and width $\ell$. (a) One-dimensional harmonic trap. (b) Two-dimensional cigar trap. In this figure, `normal junction' is the region where excess atoms do not form a ferromagnetic junction in the trap center. Parameters in each case are the same as those in Figs.\ref{fig.12} and \ref{fig.13}, respectively.}
\label{fig.14}
\end{center}    
\end{figure}

\subsection{Effects of harmonic trap}

In a trapped system, excess atoms are not always localized around the potential barrier. Indeed, as shown in Fig.\ref{fig.12} where the trap potential in Eq.(\ref{trap1D}) is used, excess atoms are only localized at the edges of the gas cloud when the central barrier is low (panels (a1) and (a2)). In this case, the potential barrier is {\it non-magnetic}, and  the $0$-phase is realized, as shown in panel (a3). As one increases the barrier height, the barrier is magnetized in the sense that an excess atoms is localized around it, leading to the $\pi$-phase (panels (b)). The number of localized atoms around the central barrier further increases, as one further increases the potential height. In panels (c), the 0-phase is again realized due to the thick ferromagnetic junction by two excess atoms. 
\par
The competition between the localization of excess atoms at the edges of the trap and magnetization of the central potential barrier can be also seen in the two-dimensional cigar trap given by Eq.(\ref{trap2D}). We show the result for this case in Fig.\ref{fig.13}. Since the present system size is rather small due to computational problems, the $\pi$-phase in panels (b) soon changes into the $0$-phase shown in panels (c), when the number of excess atoms localized around the central barrier increases only by one. However, as discussed in Sec. IV.A, this sensitivity would be suppressed when we consider a lager system.
\par
Figure \ref{fig.14} shows the phase diagram in terms of the potential height and width. In both the one- and two-dimensional cases, we obtain finite regions of $\pi$-phase. Thus, the $\pi$-phase is expected to be experimentally accessible even for a trapped superfluid Fermi gas.
\par
\section{summary}
To summarize, we have investigated the possibility of SFS-junction in a superfluid Fermi gas. Using the phase separation of a polarized Fermi gas, we showed that a non-magnetic potential barrier embedded in the system can be magnetized in the sense that some of excess atoms are localized around it. The polarized barrier works as a ferromagnetic junction, leading to the $\pi$-phase. We also discussed how the presence of harmonic trap, as well as the dimensionality of system, affect the SFS-junction and $\pi$-phase.
\par
For the observation of $\pi$-phase, since the phase of the order parameter differs by $\pi$ across the junction, an interference experiment might be useful. Since it is known that low-energy bound states may appear around the nodes of order parameter\cite{Chu,OhashiTakada}, the observation of these ingap states in single-particle excitation spectra may be another idea to detect the $\pi$-phase. 
\par
Besides the importance of the $\pi$-phase itself, it is also useful for various applications. For example, since it may be viewed as a part of FFLO state, one might use it to realize the FFLO state in cold Fermi gases. In addition, when the SFS-junction and $\pi$-phase are realized in a superfluid Fermi gas confined in a toroidal trap, the so-called $\pi$-junction SQUID\cite{Sigrist,Wollman} may be realized in a superfluid Fermi gas. Although we need further careful analyses for them, we expect that our results would be useful for the study of magnetic effects on Fermi superfluids by using cold Fermi gases.


\section{Acknowledgements}

We would like to thank S. Watabe, D. Inotani, and R. Watanabe for useful discussions. T. K. was supported by Global COE Program "High-Level Global Cooperation for Leading-Edge Platform on Access Space (C12)". Y. O. was supported by a Grant-in-Aid for Scientific research from MEXT in Japan (20500044).



\begin{thebibliography}{99}
\bibitem{note} When we take a very small $U$, the FFLO state becomes more stable than the phase separation. However, we do not consider such a case.
\bibitem{note4} In the case of Fig.\ref{fig.6} ($L_x=300$, $N=160$, and $U/t=4$), the Fermi momentum is evaluated as $p_F\simeq \pi\times (40/300)$ (for $N_\sigma=N/2=80$, ignoring the imbalance), so that we obtain $v_F=2t\sin p_F\simeq 0.8t$. The averaged polarization in the ferromagnetic junction ($70\lesssim i_x\lesssim 80$) is evaluated from Fig.\ref{fig.5}, as $\overline{s_i}\simeq 0.25$. For the chemical potential, our numerical result givens $\mu_\uparrow-\mu_\downarrow\simeq 2.1$. From these values, we obtain $\lambda\simeq 9.1$.
\end{thebibliography}
\end{document}